\documentclass[aps,prl,showpacs,twocolumn,groupedaddress,floatfix,amsfonts]{revtex4}
\usepackage{graphicx}
\usepackage{dcolumn}
\usepackage{bm}
\usepackage{amssymb}
\usepackage{amsmath}
\usepackage{xspace}

\newcommand{\triplexs}{\ensuremath{\mathrm{d}^3\sigma/(\mathrm{d}\Ptg\mathrm{d}y^\gamma \mathrm{d}y^\text{jet})}\xspace}
\newcommand{\GeV}{\ensuremath{\text{GeV}}\xspace}
\newcommand{\TeV}{\ensuremath{\text{TeV}}\xspace}
\newcommand{\Ptg}{p_{T}^{\gamma}\xspace}
\newcommand{\ptg}{\ensuremath{p_T^{\gamma}}\xspace}
\newcommand{\gb}{\ensuremath{\gamma+{b}+X}\xspace}
\newcommand{\gc}{\ensuremath{\gamma+{c}+X}\xspace}
\newcommand{\rJLIP}{\ensuremath{P_{\text{HF-jet}}}\xspace}

\begin{document}

\hspace{5.2in} \mbox{FERMILAB-PUB-08-582-E}

\title{\boldmath Measurement of $\gamma+b+X$ and $\gamma+c+X$ production 
cross sections in $p\bar{p}$ collisions at $\sqrt{s}=1.96~\TeV$}
%
\author{V.M.~Abazov$^{36}$}
\author{B.~Abbott$^{75}$}
\author{M.~Abolins$^{65}$}
\author{B.S.~Acharya$^{29}$}
\author{M.~Adams$^{51}$}
\author{T.~Adams$^{49}$}
\author{E.~Aguilo$^{6}$}
\author{M.~Ahsan$^{59}$}
\author{G.D.~Alexeev$^{36}$}
\author{G.~Alkhazov$^{40}$}
% ...

\affiliation{\vspace{0.1 in}(The D\O\ Collaboration)\vspace{0.1 in}}
% ...
\affiliation{$^{6}$University of Alberta, Edmonton, Alberta, Canada,
                Simon Fraser University, Burnaby, British Columbia, Canada,
                York University, Toronto, Ontario, Canada, and
                McGill University, Montreal, Quebec, Canada}
% ...
\affiliation{$^{29}$Tata Institute of Fundamental Research, Mumbai, India}
% ...
\affiliation{$^{36}$Joint Institute for Nuclear Research, Dubna, Russia}
% ...
\affiliation{$^{40}$Petersburg Nuclear Physics Institute,
                St. Petersburg, Russia}
% ...
\affiliation{$^{49}$Florida State University, Tallahassee, Florida 32306, USA}
% ...
\affiliation{$^{51}$University of Illinois at Chicago,
                Chicago, Illinois 60607, USA}
% ...
\affiliation{$^{59}$Kansas State University, Manhattan, Kansas 66506, USA}
% ...
\affiliation{$^{65}$Michigan State University,
                East Lansing, Michigan 48824, USA}
% ...
\affiliation{$^{75}$University of Oklahoma, Norman, Oklahoma 73019, USA}
% ...
\date{January 6, 2009}

\begin{abstract}
  First measurements of the differential cross sections \triplexs for
  the inclusive production of a photon in association with a heavy
  quark ($b$, $c$) jet are presented, covering photon transverse
  momenta $30<\Ptg< 150$~\GeV, photon rapidities $| y^\gamma| < 1.0$,
  jet rapidities $|y^\text{jet}| < 0.8$, and jet transverse momenta
  $p_T^{\rm jet}>15$~\GeV.  The results are based on an integrated
  luminosity of $1$~fb${}^{-1}$ in $p\bar{p}$ collisions at
  $\sqrt{s}=1.96~\TeV$ recorded with the D0 detector at the Fermilab
  Tevatron Collider.  The results are compared with next-to-leading
  order perturbative QCD predictions.
\end{abstract}
\maketitle


Photons ($\gamma$) produced in association with heavy quarks $Q$
($\equiv c$ or $b$) in the final state of hadron-hadron interactions
provide valuable information about the parton distributions of the
initial state hadrons~\cite{Berger_charm,CTEQ_c}.  Such events are
produced primarily through the QCD Compton-like scattering process
$gQ\to \gamma Q$, which dominates up to photon transverse momenta
($\Ptg$) of $\sim90$~GeV for \gc and up to $\sim120$~GeV for \gb
production, but also through quark-antiquark annihilation $q\bar{q}\to
\gamma g \to \gamma Q\bar{Q}$.  Consequently, $\gamma + Q+X$
production is sensitive to the $b$, $c$, and gluon ($g$) densities
within the colliding hadrons, and can provide constraints on parton
distribution functions (PDFs) that have substantial
uncertainties~\cite{WKTung,CTEQ}.  The heavy quark and gluon content
is an important aspect of QCD dynamics and of the fundamental
structure of the proton.  In particular, many searches for new
physics, e.g. for certain Higgs boson production
modes~\cite{Diff_Higgs,Ch_Higgs,Higgs_rev,HF_PDF_Gluck}, will benefit
from a more precise knowledge of the heavy quark and gluon content of
the proton.

This Letter presents the first measurements of the inclusive
differential cross sections \triplexs for $\gamma+b+X$ and
$\gamma+c+X$ production in $p\bar{p}$ collisions, where $y^\gamma$ and
$y^\text{jet}$ are the photon and jet rapidities~\cite{rapidity}.  The
results are based on an integrated luminosity of
1.02$~\pm~0.06$~fb${}^{-1}$ \cite{lumi} collected with the D0
detector~\cite{D0_detector} at the Fermilab Tevatron Collider at
$\sqrt{s}=1.96~\TeV$.  The highest $p_T$ (leading) photon and jet are
required to have $|y^{\gamma}|<1.0$ and $|y^\text{jet}|<0.8$, and
transverse momentum $30<\Ptg< 150$~GeV and $p_{T}^{\rm jet}>15$~GeV.
This selection allows one to probe PDFs in the range of
parton-momentum fractions $0.01\lesssim x\lesssim 0.3$, and hard
scatter scales of $9\times 10^2 \lesssim Q^2 \equiv
(p_{T}^{\gamma})^2\lesssim 2\times 10^4~\GeV^2$.  Differential cross
sections are presented for two regions of kinematics, defined by
$y^{\gamma} y^\text{jet}>0$ and $y^{\gamma} y^\text{jet}<0$.  These
two regions provide greater sensitivity to the parton $x$ because they
probe different sets of $x_1$ and $x_2$ intervals, as discussed in
Ref.~\cite{gamjet_PLB}.

The triggers for this analysis identify clusters of large
electromagnetic (EM) energy, and are based on $p_T^\gamma$ and on the
spatial distribution of energy in the photon shower.  The trigger
efficiency is $\approx$96\% for photon candidates with $p_T^\gamma =
30$~GeV and rises to nearly 100\% for $p_T^\gamma>40$~GeV.

To reconstruct photon candidates, towers~\cite{D0_detector} with large
depositions of energy are used as seeds to create clusters of energy
in the EM calorimeter in a cone of radius ${\cal R}=0.4$, where ${\cal
  R}\equiv\sqrt{(\Delta\eta)^2+(\Delta\phi)^2}$~\cite{etaphi}. Once an
EM energy cluster is formed, the final energy $E_\text{EM}$ is defined
by a smaller cone of ${\cal R}=0.2$.  Photon candidates are required
to be isolated within the calorimeter, and must also have $>96$\% of
their energy in its EM section.  We require the sum of the total
energy inside a cone of ${\cal R}=0.4$, after the subtraction of
$E_\text{EM}$, to be $<7$\% of~$E_\text{EM}$.  We also require the
width of the energy-weighted shower in the most finely segmented part
of the EM calorimeter to be consistent with that expected for an
electromagnetic shower, and the probability for any track spatially
matched to the photon EM cluster to be $<$0.1\%.  Background from
dijet events containing $\pi^0$ and $\eta$ mesons that can mimic
photon signatures is also rejected using an artificial neural network
for identifying photons ($\gamma$-ANN), described in Ref.~\cite{gamjet_PLB}. 
The requirement that the $\gamma$-ANN output be $>0.7$, combined with 
all other photon selection critera, reduces the dijet event efficiency 
to 0.1--0.5\%. We calculate
photon detection efficiencies using a Monte Carlo (MC) simulation.
Signal events are generated using {\sc pythia}~\cite{PYT} and
processed through a {\sc geant}-based~\cite{Geant} simulation of the
detector geometry and response, and reconstructed using the same
software as for the data.  The MC efficiencies are calibrated to those
in data using small correction factors measured in $Z\to e^+e^-$
samples.  The total efficiency of the above photon selection criteria
is 63--80\%, depending on \ptg.  The systematic uncertainties on these
values are 5\%, and are mainly due to uncertainties in the isolation,
the track-match veto, and the $\gamma$-ANN requirements.

At least one jet must be present in each event. Jets are reconstructed
using the D0 Run~II algorithm~\cite{c:Run2Cone} with a radius of
$0.5$.  The efficiency for a jet to be reconstructed and to satisfy
the jet identification criteria is 93\%, 96.5\%, and 94.5\% for light
($u$, $d$, $s$ quark or $g$), $c$, and $b$ jets at $\ptg=30~\GeV$ and
increases to $\approx 98$\% at $\ptg=150$~GeV, independent of the jet
flavor. The impact from uncertainties on jet energy scale, jet energy
resolution, and difference in energy response between light and $b(c)$
jets is found to be between 8\,\%(6\,\%) and 2\,\%(2\,\%) for
$p_T^{\rm jet}$ between 15~\GeV and 150~\GeV.  The leading jet is also
required to have at least two associated tracks with $p_T>0.5$~\GeV
and the track leading in $p_T$ must have $p_T>1.0$~\GeV, and each
track must have at least one hit in the silicon microstrip tracker.
These criteria ensure that the jet has sufficient information to be
classified as a heavy-flavor (HF) candidate.  Light jets are suppressed
using a dedicated artificial neural network ($b$-ANN)~\cite{c:bNN}
that exploits the longer lifetimes of heavy-flavor hadrons relative to
their lighter counterparts. The leading jet is required to have a
$b$-ANN output $>0.85$.  Depending on $\ptg$, this selection is
55--62\% efficient for $\gamma+b$ jet, and 11--12\% efficient for
$\gamma+c$ jet events, with 3--5\% relative uncertainties on these
values.  Only 0.2--1\% of light jets are misidentified as heavy-flavor
jets.

A primary collision vertex with $\ge$3~tracks is required within 35~cm
of the center of the detector along the beam axis.  The missing
transverse momentum in the event is required to be $<0.7
p_{T}^{\gamma}$ so as to suppress background from cosmic-ray muons and
$W\to\ell\nu$ decays.  Such a requirement is highly efficient for
signal, achieving an efficiency $\geq 96\%$ even for events with
semi-leptonic heavy-flavor quark decays.

About 13,000 events remain in the data sample after applying all 
selection criteria. Background for photons, stemming mainly from 
dijet events in which one jet is misidentified as a photon, is 
still present in this sample. To estimate the photon purity, a 
template fitting technique is employed~\cite{Templates}. The 
$\gamma$-ANN distribution in data is fitted to a linear 
combination of templates for photons and jets obtained from
simulated $\gamma~+$ jet and dijet samples, respectively.  An
independent fit is performed in each $\Ptg$ bin, yielding photon
purities between 51\% and 93\% for $30<\ptg<150~\GeV$.  The fractional
contributions of $b$ and $c$ jets are determined by fitting templates
of $\rJLIP=-\ln\prod_{i}{P_{\rm track}^{i}}$ to the data, where
$P_{\rm track}^{i}$ is the probability that a track originates from
the primary vertex, based on the significance of the track's distance
of closest approach to the primary vertex.  All tracks within the jet
cone are used in the fit, except the one with lowest value of $P_{\rm
  track}$. Jets from $b$ quarks usually have large values of \rJLIP,
whereas light jets mostly have small values, as their tracks originate
from the primary vertex.  Templates are used for the shape information
of the $\rJLIP$ distributions.  For $b$ and $c$ jets these are
extracted from MC events whereas the light jet template is taken from
a data sample enriched in light jets, which is corrected for
contributions from $b$ and $c$ quarks.
\begin{figure}
\includegraphics[width=7cm,height=6.5cm,trim=27 30 55 50,clip=true]{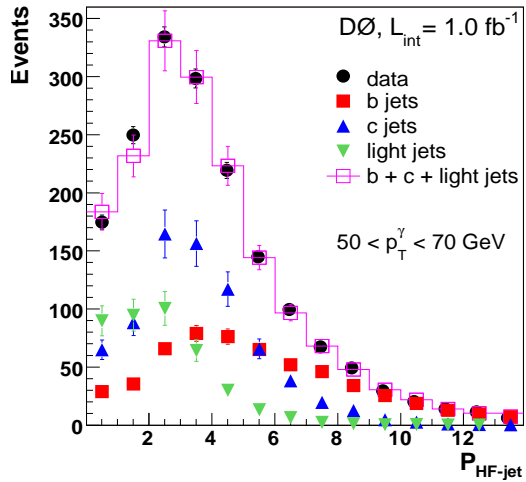}
\caption{Distribution of observed events for \rJLIP
  after all selection criteria for the bin $50<\Ptg< 70$~GeV.  The
  distributions for the $b$, $c$, and light jet templates are shown
  normalized to their fitted fraction.  Error bars on the templates
  represent combined uncertainties from statistics of the MC and the
  fitted jet flavor fractions, while the data contain just statistical
  uncertainties.  Fits in the other \ptg bins are of similar quality.}
\label{fig:cbjet_test}
\end{figure}
The result of a maximum likelihood fit, normalized to the number of
events in data, is shown in Fig.~\ref{fig:cbjet_test} for
$50<\Ptg<70~\GeV$.  The estimated fractions of $b$ and $c$ jets in all
$\Ptg$ bins vary between 25--34\% and 40--48\%, respectively.  The
corresponding uncertainties range between
7-24\%, dominated at higher $\Ptg$ by the limited data statistics.

\begin{figure}
\includegraphics[width=7cm,height=6.5cm,trim=10 20 55 50,clip=true]{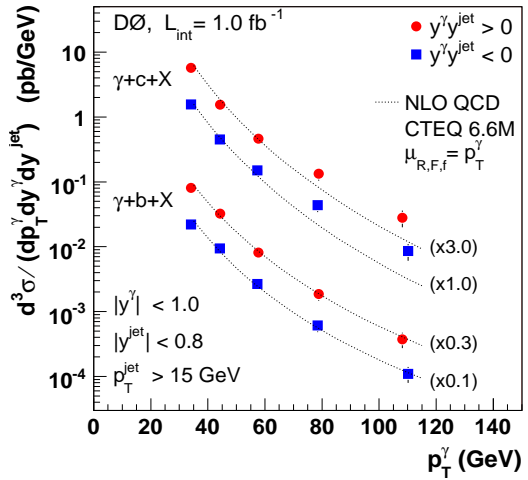}
\caption{ The \gb and \gc differential cross sections
  as a function of $\Ptg$ in the two regions $y^{\gamma}
  y^\text{jet}>0$ and $y^{\gamma} y^\text{jet}<0$.  The uncertainties
  on the data points include statistical and systematic contributions
  added in quadrature. The NLO pQCD predictions using {\sc cteq}6.6M
  PDFs are indicated by the dotted lines.}
\label{fig:xsectbc1plot}
\end{figure}

The differential cross sections are extracted in five bins of \ptg{}
and in the two regions of $y^\gamma y^\text{jet}$, and are all listed
in Table~\ref{tab:results}.  The measured cross sections are corrected
for the effect of finite calorimeter energy resolution affecting
$\Ptg$ using the unfolding procedure described in
Ref.~\cite{D0_unsmearing}. Such corrections are 1--3\%.  The measured
differential cross sections are shown in Fig.~\ref{fig:xsectbc1plot}
for \gb and \gc production as a function of $\Ptg$ for the jet and
photon rapidity intervals in question.  The cross sections fall by
more than three orders of magnitude in the range $30<\Ptg< 150~\GeV$.
The statistical uncertainty on the results ranges from 2\% in the
first $\Ptg$ bin to $\approx 9\%$ in the last bin, while the total
systematic uncertainty varies between 15\% and 28\%.  The main
uncertainty at low \ptg{} is due to the photon purity (10.5\%) and the
heavy-flavor fraction fit (9\%).  At higher \ptg{}, the uncertainty is
dominated by the heavy-flavor fraction.  Other significant
uncertainties result from the jet-selection efficiency (between 8\%
and 2\%), the photon selection efficiency (5\%), and the luminosity
(6.1\%)~\cite{lumi}.  Systematic uncertainties have a 60--68\%
correlation between adjacent $\Ptg$ bins for $30<\Ptg<50$~GeV and
20--30\% for $\Ptg>$70 GeV.

\begin{figure}
\includegraphics[width=\linewidth]{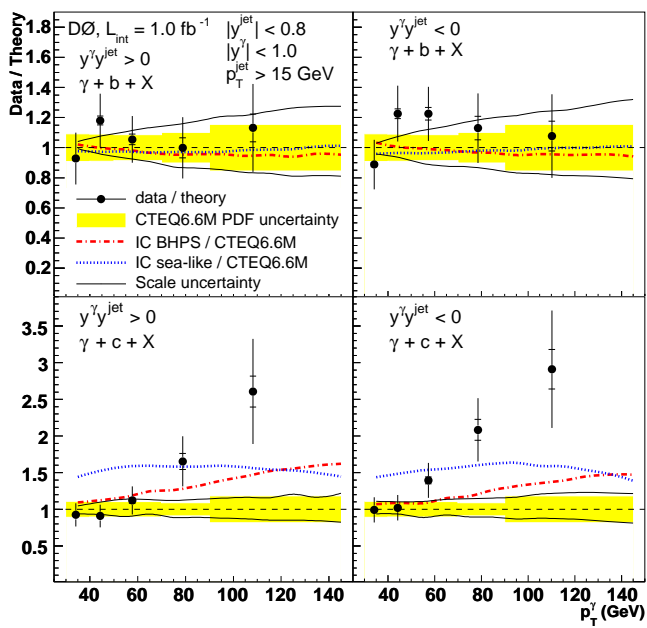}
\caption{The data-to-theory ratio of cross sections as a function of
  $\Ptg$ for \gb and \gc in the regions $y^{\gamma} y^\text{jet}>0$
  and $y^{\gamma} y^\text{jet}<0$.  The uncertainties on the data
  include both statistical (inner line) and full uncertainties (entire
  error bar).  Also shown are the uncertainties on the theoretical
  pQCD scales and the {\sc cteq}6.6M PDFs.  The scale uncertainties
  are shown as dotted lines and the PDF uncertainties by the shaded
  regions.  The ratio of the standard {\sc cteq}6.6M prediction to two
  models of intrinsic charm is also shown.}
\label{fig:xsectb1ratio}
\end{figure}

\begin{table*}
  \centering
  \caption{The \gb and \gc cross sections in bins of $\Ptg$ in the two regions $y^{\gamma}
  y^\text{jet}>0$ and $y^{\gamma} y^\text{jet}<0$ together with
 statistical, $\delta\sigma_{\text{stat}}$,  and systematic, $\delta\sigma_{\text{syst}}$, uncertainties. 
  The theory cross sections $\sigma_{\text{theory}}$ are taken from Ref.~\cite{Tzvet}.}
  \label{tab:results}
  \begin{tabular}{ccp{0.025\linewidth}cccccp{0.025\linewidth}ccccc}
    \hline
    \hline
    &&& \multicolumn{5}{c}{$y^{\gamma} y^\text{jet}>0$} && \multicolumn{5}{c}{$y^{\gamma} y^\text{jet}<0$} \\
    \hline
    &$\Ptg$ bin && $\langle\Ptg\rangle$ & Cross section &
    $\delta\sigma_{\text{stat}}$ & $\delta\sigma_{\text{syst}}$ & $\sigma_{\text{theory}}$ && $\langle\Ptg\rangle$ & Cross section & $\delta\sigma_{\text{stat}}$ & $\delta\sigma_{\text{syst}}$ & $\sigma_{\text{theory}}$\\
        & (GeV)     & & (GeV)    & (pb/GeV)             & ($\%$)&($\%$)& (pb/GeV)            && (GeV)    & (pb/GeV)              & ($\%$)&($\%$)& (pb/GeV)\\
    \hline          
    \gb & 30--40    & & 34.1     & 2.73$\times 10^{-1}$ &  1.5 & 18.5 & 2.96$\times 10^{-1}$ && 34.1     & 2.23$\times 10^{-1}$  &  1.6  & 19.1 & 2.45$\times 10^{-1}$\\
        & 40--50    & & 44.3     & 1.09$\times 10^{-1}$ &  2.5 & 15.5 & 9.31$\times 10^{-2}$ && 44.2     & 9.53$\times 10^{-2}$  &  2.6  & 16.0 & 8.18$\times 10^{-2}$\\
        & 50--70    & & 57.6     & 2.72$\times 10^{-2}$ &  3.3 & 15.2 & 2.66$\times 10^{-2}$ && 57.4     & 2.67$\times 10^{-2}$  &  3.3  & 15.3 & 2.22$\times 10^{-2}$\\
        & 70--90    & & 78.7     & 6.21$\times 10^{-3}$ &  6.6 & 20.8 & 6.39$\times 10^{-3}$ && 78.3     & 6.10$\times 10^{-3}$  &  6.7  & 20.8 & 5.49$\times 10^{-3}$\\
        & 90--150   & & 108.3    & 1.23$\times 10^{-3}$ &  8.2 & 26.2 & 1.11$\times 10^{-3}$ && 110.0    & 1.09$\times 10^{-3}$  &  8.9  & 25.7 & 1.05$\times 10^{-3}$\\
    \hline                                                                                                                                                           
    \gc & 30--40    & & 34.1     & 1.90                 &  1.5 & 18.1 & 2.02                 && 34.1     & 1.56                  &  1.6  & 18.7 & 1.59                \\
        & 40--50    & & 44.3     & 5.14$\times 10^{-1}$ &  2.5 & 17.7 & 5.82$\times 10^{-1}$ && 44.2     & 4.51$\times 10^{-1}$  &  2.6  & 18.1 & 4.56$\times 10^{-1}$\\
        & 50--70    & & 57.6     & 1.53$\times 10^{-1}$ &  3.3 & 17.9 & 1.41$\times 10^{-1}$ && 57.4     & 1.50$\times 10^{-1}$  &  3.3  & 18.0 & 1.10$\times 10^{-1}$\\
        & 70--90    & & 78.7     & 4.45$\times 10^{-2}$ &  6.6 & 21.3 & 2.85$\times 10^{-2}$ && 78.3     & 4.39$\times 10^{-2}$  &  6.7  & 21.3 & 2.22$\times 10^{-2}$\\
        & 90--150   & & 108.3    & 9.63$\times 10^{-3}$ &  8.2 & 27.5 & 3.69$\times 10^{-3}$ && 110.0    & 8.57$\times 10^{-3}$  &  8.9  & 27.0 & 3.28$\times 10^{-3}$\\
    \hline
    \hline
  \end{tabular}
\end{table*}

Next-to-leading order (NLO) perturbative QCD (pQCD) predictions, with
the renormalization scale $\mu_{R}$, factorization scale $\mu_{F}$,
and fragmentation scale $\mu_f$, all set to $\Ptg$, are also given in
Table~\ref{tab:results} and compared to data in
Fig.~\ref{fig:xsectbc1plot}.  These predictions~\cite{Tzvet} are
are based on techniques used to calculate the cross
section analytically~\cite{Harris}, and the ratios of the measured to the
predicted cross sections are shown in Fig.~\ref{fig:xsectb1ratio}.

The uncertainty from the choice of the scale is estimated through a
simultaneous variation of all three scales by a factor of two, i.e.,
to $\mu_{R,F,f}=0.5 p_T^\gamma$ and $2 p_T^\gamma$.  The predictions
utilize {\sc cteq}6.6M PDFs~\cite{CTEQ}, and are corrected for effects
of parton-to-hadron fragmentation.  This correction for $b\,(c)$ jets
varies from $7.5$\% ($3$\%) at $30<\Ptg<40~\GeV$ to 1\% at
$90<\Ptg<150~\GeV$.

The pQCD prediction agrees with the measured cross sections for \gb
production over the entire \ptg{} range, and with \gc production for
$\ptg<70$~GeV.  For $\ptg>70$~GeV, the measured \gc cross section is
higher than the prediction by about 1.6--2.2 standard deviations
(including only the experimental uncertainties) with the difference
increasing with growing \ptg.

Parameterizations for two models containing intrinsic charm (IC) have
been included in {\sc cteq}6.6 \cite{CTEQ_c}, and their ratios to the
standard {\sc cteq} predictions are also shown in
Fig.~\ref{fig:xsectb1ratio}.  Both non-perturbative models predict a
higher \gc cross section. In the case of the BHPS model~\cite{CTEQ_c}
it grows with $\Ptg$.  The observed difference may also be caused by
an underestimated contribution from the $g\to Q\bar{Q}$ splitting in
the annihilation process that dominates for $\Ptg>90~\GeV$~\cite{PDG}.

In conclusion, we have performed the first measurement of the
differential cross section of inclusive photon production in
association with heavy flavor ($b$ and $c$) jets at a $p\bar{p}$
collider. The results cover the range $30<\ptg<150~\GeV$,
$|y^\gamma|<1.0$, and $|y^{\rm jet}|<0.8$. The measured cross sections
provide information about $b$, $c$, and gluon PDFs for $0.01\lesssim
x\lesssim 0.3$.  NLO pQCD predictions using {\sc cteq}6.6M
PDFs~\cite{Tzvet} for \gb production agree with the measurements over
the entire \ptg range.  We observe disagreement between theory and
data for \gc production for $\ptg>70$~GeV.

We are very grateful to the authors of the theoretical code,
Tzvetalina Stavreva and Jeff Owens, for providing predictions and for
many fruitful discussions.  
%
We thank the staffs at Fermilab and collaborating institutions, 
and acknowledge support from the 
DOE and NSF (USA);
CEA and CNRS/IN2P3 (France);
FASI, Rosatom and RFBR (Russia);
CNPq, FAPERJ, FAPESP and FUNDUNESP (Brazil);
DAE and DST (India);
Colciencias (Colombia);
CONACyT (Mexico);
KRF and KOSEF (Korea);
CONICET and UBACyT (Argentina);
FOM (The Netherlands);
STFC (United Kingdom);
MSMT and GACR (Czech Republic);
CRC Program, CFI, NSERC and WestGrid Project (Canada);
BMBF and DFG (Germany);
SFI (Ireland);
The Swedish Research Council (Sweden);
CAS and CNSF (China);
and the
Alexander von Humboldt Foundation (Germany).

\bibliography{prl}

\begin{thebibliography}{33}
\expandafter\ifx\csname natexlab\endcsname\relax\def\natexlab#1{#1}\fi
\expandafter\ifx\csname bibnamefont\endcsname\relax
  \def\bibnamefont#1{#1}\fi
\expandafter\ifx\csname bibfnamefont\endcsname\relax
  \def\bibfnamefont#1{#1}\fi
\expandafter\ifx\csname citenamefont\endcsname\relax
  \def\citenamefont#1{#1}\fi
\expandafter\ifx\csname url\endcsname\relax
  \def\url#1{\texttt{#1}}\fi
\expandafter\ifx\csname urlprefix\endcsname\relax\def\urlprefix{URL }\fi
\providecommand{\bibinfo}[2]{#2}
\providecommand{\eprint}[2][]{\url{#2}}

\bibitem[{\citenamefont{Aurenche {\sl et~al.}}(1996)}]{Berger_charm}
\bibinfo{author}{\bibfnamefont{B.}~\bibnamefont{Bailey}},
\bibinfo{author}{\bibfnamefont{E.L.}~\bibnamefont{Berger}},
\bibinfo{author}{\bibfnamefont{L.E.}~\bibnamefont{Gordon}},
\bibinfo{journal}{Phys. Rev.}
{\bibinfo{volume}D {\bf 54}}, \bibinfo{pages}{1896} (\bibinfo{year}{1996}).

\bibitem[{\citenamefont{Pumplin {\sl et~al.}}(2007)}]{CTEQ_c}
\bibinfo{author}{\bibfnamefont{J.}~\bibnamefont{Pumplin}}, 
\bibinfo{author}{\bibfnamefont{H.L.}~\bibnamefont{Lai}}, 
\bibinfo{author}{\bibfnamefont{W.K.}~\bibnamefont{Tung}}, 
\bibinfo{journal}{Phys. Rev.}
{\bibinfo{volume}D {\bf 75}}, \bibinfo{pages}{054029} (\bibinfo{year}{2007}).

\bibitem[{\citenamefont{Wu-Ki {\sl et~al.}}(2005)}]{WKTung}
  \bibinfo{author}{\bibfnamefont{W.K.}~\bibnamefont{Tung}}, 
  \eprint{arXiv:hep-ph/0409145} (\bibinfo{year}{2004}).
  
\bibitem[{\citenamefont{D. Stump {\sl et~al.}}(2003)}]{CTEQ}
  \bibinfo{author}{\bibfnamefont{D.}~\bibnamefont{Stump}} \bibnamefont{{\sl et~al.}},
  \bibinfo{journal}{JHEP} \textbf{\bibinfo{volume}{0310}}, \bibinfo{pages}{046}
  (\bibinfo{year}{2003}).
  
\bibitem[{\citenamefont{Brodsky {\sl et~al.}}(2006)}]{Diff_Higgs}

\bibinfo{author}{\bibfnamefont{S.J.}~\bibnamefont{Brodsky}},
\bibinfo{author}{\bibfnamefont{B.}~\bibnamefont{Kopeliovich}},
\bibinfo{author}{\bibfnamefont{I.}~\bibnamefont{Schmidt}},
\bibinfo{author}{\bibfnamefont{J.}~\bibnamefont{Soffer}},
\bibinfo{journal}{Phys. Rev.}
{\bibinfo{volume}D {\bf 73}}, \bibinfo{pages}{113005} (\bibinfo{year}{2006}).

\bibitem[{\citenamefont{He {\sl et~al.}}(2006)}]{Ch_Higgs}
  \bibinfo{author}{\bibfnamefont{H.J.}~\bibnamefont{He}},
  \bibinfo{author}{\bibfnamefont{C.P.}~\bibnamefont{Yuan}},
  \bibinfo{journal}{Phys. Rev. Lett.} \textbf{\bibinfo{volume}{83}},
  \bibinfo{pages}{28} (\bibinfo{year}{1999});
  \bibinfo{author}{\bibfnamefont{C.}~\bibnamefont{Balazs}},
  \bibinfo{author}{\bibfnamefont{H.J.}~\bibnamefont{He}},
  \bibinfo{author}{\bibfnamefont{C.P.}~\bibnamefont{Yuan}},
  \bibinfo{journal}{Phys. Rev.}
  {\bibinfo{volume}D {\bf 60}}, \bibinfo{pages}{114001} (\bibinfo{year}{1999}).
  
  
\bibitem[{\citenamefont{K.A.Assamagan {\sl et~al.}}(2003)}]{Higgs_rev}
  \bibinfo{author}{\bibfnamefont{K.A.}~\bibnamefont{Assamagan}}, 
  \eprint{arXiv:hep-ph/0406152} (\bibinfo{year}{2003}).
  
\bibitem[{\citenamefont{Gluck {\sl et~al.}}(2008)}]{HF_PDF_Gluck}
  \bibinfo{author}{\bibfnamefont{M.~} \bibnamefont{Gl\"{u}ck}} \bibnamefont{{\sl et~al.},}
  \bibinfo{journal}{Phys. Lett.} {\bibinfo{volume}B {\bf 664}},
  \bibinfo{pages}{133} (\bibinfo{year}{2008}).
  
\bibitem{rapidity}
  \bibfnamefont{Rapidity is defined as $y=-\ln[(E + p_Z)/(E - p_Z)]$, where
    $E$ is the energy and $p_Z$ is the momentum component along
    the proton beam direction.}
  
  
\bibitem[{\citenamefont{Andeen {\sl et~al.}}(1994)}]{lumi}
  \bibinfo{author}{\bibfnamefont{T.} \bibnamefont{Andeen}}
  \bibnamefont{{\sl et~al.}},
  \bibfnamefont{FERMILAB-TM-2365} (\bibinfo{year}{2007}).
  
  
\bibitem[{\citenamefont{Abazov {\sl et~al.}}(2006)}]{D0_detector}
  \bibinfo{author}{\bibfnamefont{V.M.}~ \bibnamefont{Abazov}}
  \bibnamefont{{\sl et~al.}} (\bibinfo{collaboration}{D0 Collaboration}),
  \bibinfo{journal}{Nucl. Instrum. Methods Phys. Res.} {\bibinfo{volume}A {\bf 565}}, \bibinfo{pages}{463}
  (\bibinfo{year}{2006}).
  
  
\bibitem[{\citenamefont{Abazov {\sl et~al.}}(2008)}]{gamjet_PLB}
  \bibinfo{author}{\bibfnamefont{V.M.~} \bibnamefont{Abazov}}
  \bibnamefont{{\sl et~al.}} (\bibinfo{collaboration}{D0 Collaboration}),
  \bibinfo{journal}{Phys. Lett.} {\bibinfo{volume}B {\bf 666}},
  \bibinfo{pages}{435} (\bibinfo{year}{2008}).
  
  
\bibitem{etaphi} \bibfnamefont{Pseudorapidity $\eta$ is defined as
    $\eta=-\ln[\tan(\theta/2)]$, where $\theta$ is the polar angle
    with respect to the proton beam direction, with origin at the
    center of the detector. $\phi$ is defined as the azimuthal angle
    in the plane transverse to the proton beam direction.}
  
\bibitem[{\citenamefont{Sj{\"o}strand {\sl et~al.}}(2001)}]{PYT}
  \bibinfo{author}{\bibfnamefont{T.}~\bibnamefont{Sj{\"o}strand}}
  \bibnamefont{{\sl et~al.}}, \bibinfo{journal}{Comput. Phys. Commun.}
  \textbf{\bibinfo{volume}{135}}, \bibinfo{pages}{238} (\bibinfo{year}{2001}).
  
\bibitem[{\citenamefont{Brun and Carminati}(2001)}]{Geant}
  \bibinfo{author}{\bibfnamefont{R.} \bibnamefont{Brun}} \bibnamefont{and}
  \bibinfo{author}{\bibfnamefont{F.} \bibnamefont{Carminati}},
  \bibinfo{journal}{CERN Program Library Long Writeup} \textbf{\bibinfo{volume}{W5013}},
  (\bibinfo{year}{1993}), \bibnamefont{unpublished}.
  
\bibitem[{\citenamefont{Zeppenfeld {\sl et~al.}}(1994)}]{c:Run2Cone}
  \bibinfo{author}{\bibfnamefont{G.C.}~\bibnamefont{Blazey}} \bibnamefont{{\sl et~al.}},
  \bibfnamefont{arXiv:hep-ex/0005012} (\bibinfo{year}{2000}).
  
\bibitem{c:bNN}
  \bibinfo{author}{\bibfnamefont{T.}~\bibnamefont{Scanlon}},
  \bibfnamefont{Ph.D. thesis, FERMILAB-THESIS-2006-43.}
  
\bibitem[{\citenamefont{Barlow {\sl et~al.}}(1993)}]{Templates}
  \bibinfo{author}{\bibfnamefont{R.}~\bibnamefont{Barlow}},
  \bibinfo{author}{\bibfnamefont{C.}~\bibnamefont{Beeston}},
  \bibinfo{journal}{Comput. Phys. Commun.}
  \textbf{\bibinfo{volume}{77}}, 
  \bibinfo{pages}{219} (\bibinfo{year}{1993}).

\bibitem[{\citenamefont{Stavreva {\sl et~al.}}(2003)}]{Tzvet}
  \bibinfo{author}{\bibfnamefont{T.}~\bibnamefont{Stavreva}}, 
  \bibinfo{author}{\bibfnamefont{J.F.}~\bibnamefont{Owens}}, 
  \eprint{arXiv:0901.3791v1} (\bibinfo{year}{2009}).

\bibitem[{\citenamefont{Abbott {\sl et~al.}}(2001)}]{D0_unsmearing}
  \bibinfo{author}{\bibfnamefont{B.~} \bibnamefont{Abbott}}
  \bibnamefont{{\sl et~al.}} (\bibinfo{collaboration}{D0 Collaboration}),
  \bibinfo{journal}{Phys. Rev.}
  {\bibinfo{volume}D {\bf 64}}, \bibinfo{pages}{032003} (\bibinfo{year}{2001}).
  
\bibitem[{\citenamefont{Harris {\sl et~al.}}(2002)}]{Harris}
\bibinfo{author}{\bibfnamefont{B.W.~} \bibnamefont{Harris}}, 
\bibinfo{author}{\bibfnamefont{J.F.~} \bibnamefont{Owens}}, 
\bibinfo{journal}{Phys. Rev.}
{\bibinfo{volume}D {\bf 65}}, \bibinfo{pages}{094032} (\bibinfo{year}{2002}).
  
\bibitem[{\citenamefont{Amsler {\sl et~al.}}(2008)}]{PDG}
  \bibinfo{author}{\bibfnamefont{C.~} \bibnamefont{Amsler}},
  \bibinfo{journal}{Phys. Lett.} {\bibinfo{volume}B {\bf 667}},
  \bibinfo{pages}{1} (\bibinfo{year}{2008}).
  
\end{thebibliography}


\begin{thebibliography}{99}
%
\bibitem[a]{alton}
Visitor from Augustana College, Sioux Falls, SD, USA.
\bibitem[b]{askew,gershtein}
Visitor from Rutgers University, Piscataway, NJ, USA.
\bibitem[c]{burdin}
Visitor from The University of Liverpool, Liverpool, UK.
\bibitem[d]{hensel,meyer,park,quadt}
Visitor from II. Physikalisches Institut, Georg-August-University,
  G{\"o}ttingen, Germany.
\bibitem[e]{luna-garcia}
Visitor from Centro de Investigacion en Computacion - IPN,
  Mexico City, Mexico.
\bibitem[f]{podesta-lerma}
Visitor from ECFM, Universidad Autonoma de Sinaloa, Culiac\'an, Mexico.
\bibitem[g]{voutilainen}
Visitor from Helsinki Institute of Physics, Helsinki, Finland.
\bibitem[h]{weber}
Visitor from Universit{\"a}t Bern, Bern, Switzerland.
\bibitem[i]{wenger}
Visitor from Universit{\"a}t Z{\"u}rich, Z{\"u}rich, Switzerland.
\bibitem[\ddag]{deceased}
Deceased.

%
\vskip 0.25cm

\end{thebibliography}
\bibliographystyle{apsrev}

\end{document}